# Microwave near-field helicity and its role in the matter-field interaction

E.O. Kamenetskii, R. Joffe, and R. Shavit

Department of Electrical and Computer Engineering,
Ben Gurion University of the Negev, Beer Sheva, Israel

November 16, 2011

**Abstract**

In the preceding paper, we have shown analytically that in a source-free subwavelength region of microwave fields there exist the field structures with local coupling between the time-varying electric and magnetic fields differing from the electric-magnetic coupling in regular free-space electromagnetic waves. As a source of such near fields, there is a small quasi-2D ferrite disk with the magnetic-dipolar-mode (MDM) spectra. The near fields originated from a MDM ferrite particle are characterized by topologically distinctive structures of power-flow vortices, non-zero helicity, and a torsion degree of freedom. In this paper, we present numerical and experimental studies on the microwave near-field helicity and its role in the matter-field interaction. We show that one can distinguish different microwave near-field-helicity parameters for different permittivities of dielectric samples loading a ferrite-disk sensor. We analyze a role of topological structures of the fields on the helicity properties. We demonstrate dependence of the MDM spectra and the near-field-helicity parameters from the enantiomeric properties of the loading samples.

## I. Introduction

The well known possibility of electrostatic (plasmon) resonances in metallic nanoparticles to exhibit strong localization of electric fields in a subwavelength region was used recently for ultrasensitive characterization of biomolecules [1]. In these experiments, the enhanced sensitivity of a chiroptical measurement was obtained owing to the chiral asymmetry of the metallic structures. Such a special geometry of nanoparticles can produce so-called superchiral electromagnetic fields which have a high rate of optical excitation of small chiral molecules [2].

The results shown in Refs. [1, 2] are related to optical studies and can hardly be applied to microwave frequencies, since plasmonic oscillations in metallic structures are ineffective at microwaves. There is, however, another way for subwavelength localization and chiral asymmetry of the fields in microwaves. Recent studies of interaction between microwave electromagnetic fields and small ferrite particles with magnetic-dipolar-mode (MDM) oscillations evidently demonstrate that such particles are origin of topological singularities of the microwave near fields, which are characterized by Poynting vector vortices and symmetry breakings [3]. In contrast to small plasmon-resonance particles with the electric-field localization, in small ferrite particles with MDM resonances one has subwavelength localization of both the electric and magnetic fields. At the resonance frequencies, power flows of propagating electromagnetic waves are strongly attracted by the MDM vortices at their vicinities. Coupled states of electromagnetic fields with vortices of MDM oscillations – the MDM-vortex polaritons – are physical entities with symmetry properties distinguished from symmetry properties of the regular electromagnetic near-field configurations [4]. In Ref. [5], it has been shown analytically that microwave near fields of a MDM ferrite disk are characterized by the helicity factor and a torsion degree of freedom. These symmetry properties of the microwave near fields could appear as very intriguing factors in general problems of the field-matter interaction. With use of the MDM near-field structures one acquires an effective instrument for local microwave characterization of properties of matter. In microwave experiments [6], it was shown, for the first time, that there is a transformation of the MDM spectrum due to dielectric samples abutting to the surface of a ferrite disk. Such a dielectric loading results in shifts of the MDM resonance peaks without evident destroying the shape of the spectrum peaks. The value of the peak shifts depends on the permittivity of dielectric samples [6]. Use of subwavelength microwave fields with localization and chiral asymmetry opens a perspective for unique applications. Because of the helicity structure of the MDM near-fields, one can predict carrying out a precise spectroscopic analysis of natural and artificial chiral structures at microwaves. This will pave, in partricular, the way to creating pure microwave devices for separation of biological and drug enantiomers. This also may give an answer to a controversial issue whether or not microwave irradiation can exert a non thermal effect on biomolecules [7]. In a view of these discussions, it is worth noting that in biological structures, microwave radiation can excite certain rotational transitions, which cannot be explained as classical heating effects [7]. One can predict that rotating microwave near fields originated from MDM ferrite particles [3, 4] can be used to explain such extraordinary effects in biological structures.

In this paper, we analyze numerically the symmetry properties of microwave subwavelength fields originated from a quasi-2D ferrite disk with MDM oscillations. Based on a commercial finite-element based electromagnetic solver (HFSS, Ansoft), we confirm the helicity properties of the MDM near fields predicted analytically in Ref. [5]. We present results of numerical and experimental studies of interaction of the microwave near field with the helicity properties with dielectrics characterized by different permittivity parameters. We demonstrate dependence of the near field helicity parameters on the enantiomeric properties of the loading samples. We discuss possibility for implementation of the shown effects in near field microwave microscopy for sensitive characterization of materials with chiral enantiomeric properties.

**II. Helicity density of the MDM near fields**

In the preceding paper [5] we have shown analytically that in a near-field region abutting to a quasi-2D ferrite disk with the MDM oscillations, there should exist a non-zero parameter which we call as the helicity density of the MDM fields. For a real electric field, this parameter is written as (contrarily to Ref. [5] we use here the SI system of units):

$$F \equiv \frac{\varepsilon_0}{2} \vec{E} \cdot \nabla \times \vec{E} \ . \tag{1}$$

For time-harmonic fields, we can calculate this parameter as

$$F = \frac{\varepsilon_0}{4} \text{Im} \left\{ \vec{E} \cdot \left( \vec{\nabla} \times \vec{E} \right)^* \right\} . \tag{2}$$

We can also calculate a space angle between the vectors $\vec{E}$ and $\vec{\nabla} \times \vec{E}$ as

$$\cos\alpha = \frac{\text{Im}\left\{\vec{E}\cdot\left(\vec{\nabla}\times\vec{E}\right)^*\right\}}{\left|\vec{E}\right|\left|\vec{\nabla}\times\vec{E}\right|}. \tag{3}$$

Distinct parameters of helicity density of the MDM fields become evident from numerical studies based on a commercial finite-element electromagnetic solver (HFSS, Ansoft). For a numerical analysis in the present paper, we use the same disk parameters as in Refs. [3, 4, 8, 9]: the yttrium iron garnet (YIG) disk has a diameter of $D = 3$ mm and the disk thickness is $t = 0.05$ mm; the disk is normally magnetized by a bias magnetic field $H_0 = 4900$ Oe; the saturation magnetization of the ferrite is $4\pi M_s = 1880$ G. Our analysis starts with a structure where a ferrite disk is placed inside a $TE_{10}$-mode rectangular X-band waveguide symmetrically to its walls so that the disk axis is perpendicular to a wide wall of a waveguide. The waveguide walls are made of a perfect electric conductor (PEC). For better understanding the field structures we use a ferrite disk with very small losses: the linewidth of a ferrite is $\Delta H = 0.1$ Oe. Fig. 1 shows the module of the reflection (the $S_{11}$ scattering-matrix parameter) coefficient. The resonance modes are designated in succession by numbers $n = 1, 2, 3\ldots$ An insert in Fig. 1 shows geometry of the structure. One can clearly see that, starting from the second mode, the coupled states of the electromagnetic fields with MDM vortices demonstrate split-resonance states. The properties of these coalescent resonances, denoted in Fig. 1 by single and double primes, were analyzed in details in Ref. [4].

In Refs. [3, 4], we studied general properties of the microwave near fields originated from a ferrite disk with MDM resonances. Here we calculate numerically the helicity parameter of these near fields based on Eqs. (2) and (3). For demonstration of the helicity parameter distribution in the near-field regions of a ferrite disk, we use a cross-section plane which passes through the diameter and the axis of the disk. Fig. 2 shows the helicity density parameter $F$, calculated for the 1st MDM based on numerically obtained electric near fields. For the 2nd mode (the resonance 2"), parameter $F$ is shown in Fig. 3. For the resonances denoted by single primes (the resonances 2′, 3′, etc.) in Fig. 1, as well for non-resonance frequencies, one has zero parameter $F$ (see Fig. 4). As it follows from Figs. 2 and 3, there are regions with positive and negative quantities of the helicity parameter.

As we showed in Ref. [5], the helicity properties of the near fields are strongly related to the magnetization distribution of MDMs. The present studies of the helicity factor $F$ give evidence for the relations between distributions of magnetization inside a ferrite disk and the helicity properties of the fields outside a disk. Fig. 5 (*a*), (*b*) show the numerical solutions for magnetization for the 1st and 2nd (the resonance 2") MDMs. These numerical results are in a good correspondence with the analytical solutions shown in Fig. 5 (*c*), (*d*). In the pictures in Fig. 5, we marked specific topological regions – the vortices – of magnetization. At the time variation, these vortices of magnetization, characterizing by positive and negative topological charges, rotate about a disk axis. There can be clockwise or counterclockwise rotation, depending on a direction of a bias magnetic field. One can see that while for the 1st MDM, the vortices of magnetization exist in peripheral regions of a ferrite disk, for the 2nd MDM, these vortices of magnetization are shifted to a center of a disk. In Figs. 2 and 3, we marked the positions of these magnetization vortices. It is worth noting that the magnetization vortices are situated near the regions where parameter $F$ is zero. A maximum of parameter $F$ is at a center of a ferrite disk, between the magnetization vortices. It is also worth noting that the regions where parameter $F$ is maximal correspond to the regions where there is a maximal electric field of MDM oscillations [3, 4, 8, 9].

In paper [5], we noticed that the electric fields originated from the MDMs cannot by related to the electric-polarization effects both inside a ferrite and in abutting dielectrics outside a ferrite. However, when the dielectrics are polarized by the external (DC or RF) electric fields, one should observe the influence of the dielectric properties on the oscillation spectra due to the "spin" and "orbital" angular momentums of the MDM electric fields. Based on numerical studies, in the present paper, we verify this statement and show what is a role of dielectric loadings in the spectral characteristics and helicity properties of the MDM fields. We start with a structure of a ferrite disk loading symmetrically with two dielectric cylinders. The structure is placed inside a rectangular waveguide (Fig. 6). The dielectric cylinders (every cylinder is with the diameter of 3 mm and the height of 2 mm) are electrically polarized by the RF electric field of the $TE_{10}$ mode propagating in a waveguide. The frequency characteristics of a module of the reflection coefficient for the 1st MDM at different dielectric parameters of the cylinders are shown in Fig. 7. One can see that at dielectric loadings there are coalescent resonances (the resonances 1′ and 1″). Fig. 8 shows the Poynting vector distributions above a ferrite disk at the frequencies related to the resonance 1 of an unloaded (without dielectric cylinders) ferrite disk and the resonances 1″ of a ferrite disk with dielectric loadings. These pictures, corresponding, evidently, to the known Poynting vector distributions of the 1st MDM [3, 4, 8, 9], clearly show that dielectric loadings do not destroy the entire MDM spectrum, but cause, however, the frequency shifts of the resonance peaks.

One of the main features of the frequency characteristics of a structure with the symmetrical dielectric loadings, shown in Fig. 7, is the fact that the resonances of the 1st MDM become shifted not only to the lower frequencies, but appear to the left of the Larmor frequency of an unloaded ferrite disk. For a normally magnetized ferrite disk with the pointed above quantities of the bias magnetic field and the saturation magnetization, this Larmor frequency (calculated as $f_H = \frac{1}{2\pi}\gamma H_i$, where $\gamma$ is the gyromagnetic ratio and $H_i$ is the internal DC magnetic field) is equal to $f_H = 8,456$ GHz. When a ferrite disk is without dielectric loadings, the entire spectrum of MDM oscillations is situated to the right of the Larmor frequency $f_H$. Since dielectrics do not destroy the entire MDM spectrum, one can suppose that the Larmor frequency of a structure with dielectric loadings is lower than the Larmor frequency of an unloaded ferrite disk. We clarify this statement based on the following analysis.

The electric field of MDMs has both the "spin" and "orbital" angular momentums. As we discussed in Ref. [5], the electric fields originated from a MDM ferrite disk cannot cause the electric polarization of a dielectric material. However, due to the MDM electric fields one can observe the mechanical torque exerted on a given electric dipole. This mechanical torque is defined as a cross product of the MDM electric field $\vec{E}$ and the electric moment of the dipole $\vec{p}$ [5]:

$$\vec{N} = \vec{p} \times \vec{E} .\qquad(4)$$

The dipole $\vec{p}$ appears because of the electric polarization of a dielectric by the RF electric field of a microwave system (the electric field of the $TE_{10}$ mode in a rectangular waveguide, in particular). This dipole is perpendicular to a ferrite disk. At the same time, as it was shown in Ref. [5], the MDM electric field $\vec{E}$ is the field precessing in a plane parallel to a ferrite-disk plane. When a dielectric sample is placed above a ferrite disk, the electric field $\vec{E}$ affects on the elementary dipole moments of a dielectric, causing precession of these dipole momenta. The torque exerting on the electric polarization due to the MDM electric field should be equal to reaction torque exerting on the magnetization in a ferrite disk. Because of this reaction torque, the precessing magnetic moment density of the ferromagnet will be under additional mechanical rotation at a certain frequency $\Omega$. For the magnetic moment density of the ferromagnet, $\vec{M}$, the motion equation acquires the following form (see Appendix B in Ref. [5]):

$$\frac{d\vec{M}}{dt} = -\gamma \, \vec{M} \times \left( \vec{H} - \frac{\Omega}{\gamma} \right), \tag{5}$$

The frequency $\Omega$ is defined based on both, "spin" and "orbital", momentums of the fields of MDM oscillations. One can see that at the dielectric loadings, the magnetization motion in a ferrite disk is characterized by an effective magnetic field

$$\vec{H}_{eff} = \vec{H} - \frac{\Omega}{\gamma}. \tag{6}$$

So, the Larmor frequency of a structure with dielectric loadings is at lower frequencies in comparison with such a frequency in an unloaded ferrite disk. When we put a dielectric loading above or (and) below a ferrite disk and apply to this structure an external electric field oriented along a disk axis, we have two (or three) capacitances connected in series. The capacitance of a thin-film ferrite disk is much bigger than the capacitances of dielectric samples. Thus, the surface electric charges on ferrite-disk planes will be mainly defined by the permittivity and geometry of dielectric samples. As a result, one has the MDM spectrum transformation dependable on parameters of the dielectric samples.

The effective helicity of the fields in a dielectric is due to the helicity of the MDM near field in vacuum and associated motion of the electric polarization. The recoil torque of the fields in a dielectric loading on a ferrite disk should be directly proportional to the refractive index of the dielectric. The helicity densities in dielectrics were calculated numerically based on Eq. (2). In Fig. 9, we can see the helicity density distributions in the cross-section plane passed through the diameter and the axis of the disk for different dielectric parameters of cylinders. The frequencies correspond to the 1$^{st}$ MDM. In Fig. 10 one can see the helicity parameter distributions when the cross-section planes are parallel to the ferrite-disk plane and are placed at different distances from the ferrite surface. Since amplitude of the electric field strongly reduces with increase of a distance from the ferrite surface, it will be more interesting to analyze the properties of the near-field structures when the helicity factor is normalized on the field amplitudes. The normalized helicity factor, represented as the cosine of a space angle between vectors $\vec{E}$ and $\vec{\nabla} \times \vec{E}$, was calculated numerically based on Eq. (3) and is shown in Fig. 11 for a cross-section plane passing through the diameter and the axis of a ferrite disk.

As one can see from Figs. 9 – 11, the dielectric loadings not only reduce the quantity of the helicity factor, but result in strong modification of the near-field structure. A special attention should be paid to a case of a sufficiently big dielectric constant of the loading material ($\varepsilon_r = 50$, in our case). In this case, there are singular points on an axis of a structure [points *A* and *B* in Fig. 11 (*c*)] where the helicity of the near fields changes its sign. To clarify these properties more in details, we illustrate in Fig. 12 the electric and magnetic fields in a structure of a ferrite disk with loading dielectrics ($\varepsilon_r = 50$). These field distributions, shown at a certain time phase, are characterized by "spin" and "orbital" rotations. For a comparative analysis, we illustrate initially the field structures inside a ferrite without loading dielectrics [Fig. 12 (*a*)] and inside a ferrite with loading dielectrics [Fig. 12 (*b*)]. There are evident differences between these field structures. Figs. 12 (*c*) and (*d*) show the field structures in a dielectric above a ferrite disk. As we can see in Fig. 12 (*c*), in a central region of a cutting plane placed below a singular point *A*, a space angle between the electric and magnetic fields is about $180°$. This gives a negative helicity factor. When, however, a cutting plane is placed above a singular point *A* [Fig. 12 (*d*)], a space angle between the electric and magnetic fields in a central region is about $0°$. This results in a positive helicity factor. Such specific properties of the helicity distributions, observed in a dielectric cylinder with a sufficiently big dielectric constant, can be explained from the fact that, because of the depolarization field, the electric polarization inside such a dielectric cylinder will be strongly nonhomogeneous. So, precession of the electric polarization (due to the MDM electric field $\vec{E}$) will be nonhomogeneous as well. This results in peculiar distribution of phases of the fields along an axis of a system.

As one can see from Figs. 2, 3, 9, 11, for the patterns symmetrical with respect to a normal axis, there are antisymmetrical distributions of the helicity. For the near field structure, a plane of a thin-film ferrite disk (where the electric field changes its direction [8], [9]) is a plane of a twist. The uniqueness of the observed properties of the fields in dielectric cylinders (with $\varepsilon_r = 50$) is the presence of additional planes (which pass through points *A* and *B*, above and below a ferrite disk) of a twist for the near fields. In these panes the magnetic field changes its direction. It becomes evident that an entire structure of the near fields shows the torsion degree of freedom. It is also worth noting here that in Fig. 11(*a*), a picture of the normalized helicity factor has a slight inclination from a normal axis. Such an inclination (well observed, when no dielectric loading exist) can appear as a result of an interaction of two electromagnetic-wave power flows: one is due to propagation of a waveguide mode and another – due to the near-field power-flow vortex of the MDM. When one changes the ports of a waveguide, a picture of the normalized helicity factor inclines from a normal axis to an opposite site.

When a sample of a ferrite disk with a dielectric loading is nonsymmetric with respect to a normal axis, a structure of helicity of the near field becomes nonsymmetrical as well. Such a sample, placed inside a $TE_{10}$-mode rectangular X-band waveguide, is shown in Fig. 13. The frequency characteristics of a module of the reflection coefficient for the 1$^{st}$ MDM at different parameters of a nonsymmetrical dielectric loading are shown in Fig. 14. The helicity density distributions [calculated numerically based on Eq. (2)] are represented in Fig. 15 for the cross-section plane passed through the diameter and the axis of the disk. It is evident that a loading by one dielectric cylinder results in a nonsymmetrical helicity density of the near fields. It is also evident that for different dielectric parameters of a loading cylinder, the picture of the helicity distribution inside a dielectric is strongly different from to the above results for a sample with symmetrical dielectric loadings. A more detailed analysis of the helicity properties of the near fields in a nonsymmetrical ferrite/dielectric pattern is beyond a frame of the present paper.

### III. The near-field characterization of dielectric parameters and enantiomeric properties of matter

The shown helicity parameters of the near fields reflect exclusive properties of the MDM oscillations, which can find applications for the near-field characterization of materials at microwaves. For this purpose, it is more preferable to use an open-access microstrip structure with a ferrite-disk sensor, instead of a closed waveguide structure studied above. Fig. 16 represents the frequency characteristic of a module of the transmission (the $S_{21}$ scattering-matrix parameter) coefficient for a microstrip structure with a thin-film ferrite disk. Geometry of a structure is shown in an insert. In a discussed above waveguide structure with an enclosed ferrite disk, the main features of the MDM spectra are evident from the reflection characteristics. Contrarily, in the shown microstrip structure, the most interesting are the transmission characteristics. In Fig. 16, one can see the 1$^{st}$ and 2$^{nd}$ resonances of the radial variation. The frequencies of these eigenmode resonances are in a good correspondence with the frequencies of the 1$^{st}$ and 2$^{nd}$ resonances shown in Fig. 1 for a waveguide structure. Between the 1$^{st}$ and 2$^{nd}$ resonances of the radial variations, in Fig. 16 one can see the resonance of the azimuth mode. This resonance appears because of the azimuth nonhomogeneity of a microstrip structure. A detailed classification of the radial- and azimuth-variation MDMs in a ferrite disk can be found from the analytical studies in Ref. [10]. The helicity density distribution for the 1$^{st}$ radial-variation resonance in a microstrip structure with a ferrite disk is shown in Fig. 17 (*a*). Due to a metallic ground plane and dielectric properties of a substrate in a microstrip structure, there is slight nonsymmetry of the helicity distribution with respect to a disk plane. It is worth noting that the helicity properties of the near fields appear only at the MDM resonances. At non-resonance frequencies, there is zero helicity of the field structure [see Fig. 17 (*b*)].

Fig. 18 shows the spectrum transformation due to a dielectric placed above a ferrite disk. In there numerical studies of a loading of a ferrite disk in a microstrip structure, we used dielectric cylinders with the same parameters as above: the diameter of 3mm and the height of 2 mm; the dielectric constants of $\varepsilon_r = 30$ and $\varepsilon_r = 50$. We can see shifts of the peaks depending on the dielectric properties of a sample. It is evident that for the 1$^{st}$ radial-variation MDM resonance in a microstrip structure, there is a sufficiently good correspondence of the peak positions with the results in a waveguide structure shown in Fig. 14.

Our numerical studies of the microwave field helicity and its role in the matter-field interaction open a perspective for the experimental near-field characterization of material parameters. An experimental microstrip structure is realized on a dielectric substrate (Taconic RF-35, $\varepsilon_r = 3.52$, thickness of 1.52 mm). Characteristic impedance of a microstrip line is 50 Ohm. For dielectric loadings, we used cylinders of commercial microwave dielectric (non magnetic) materials with the dielectric permittivity parameters of $\varepsilon_r = 30$ (K-30; TCI Ceramics Inc) and $\varepsilon_r = 50$ (K-50; TCI Ceramics Inc). The experimental results for characterization of dielectric properties of materials are shown in Fig. 19. One can see a sufficiently good correspondence between the numerical and experimental results. It is necessary to note here that instead of a bias magnetic field in numerical studies ($H_0 = 4900$ Oe), in the experiments we applied lower quantity of a bias magnetic field: $H_0 = 4708$ Oe. Use of such a lower quantity (giving us the same positions of the non-loading-ferrite resonance peak in the numerical studies and in the experiments) is necessary because of non-homogeneity of an internal DC magnetic field in real ferrite disks. A more detailed discussion on a role of non-homogeneity of an internal DC magnetic field in the MDM spectral characteristics can be found in Refs. [10, 11].

With use of the MDM near-field structures one may acquire an effective instrument for local characterization of special topological properties of matter. This, in particular, will allow realization of microwave devices for precise spectroscopic analysis of materials with chiral structures such, for example, as biological and drug enantiomers. In Ref. [5] we showed analytically that the helicity density *F* above and below a ferrite disk should have different signs for different orientations of a normal bias magnetic field. The present numerical results confirm this prediction. Fig. 20 shows the *F*-factor distributions for a ferrite disk (without a loading dielectric) in a microstrip structure at opposite directions of a normal bias magnetic field [in a case of a ferrite disk in a waveguide, one has the same change of a sign of parameter *F* when a direction of a normal bias magnetic field changes]. This confirms that the near-field structure of the MDM electric field is characterized by the space and time symmetry breakings. Based on numerical studies, we show how these properties of the MDM near fields can be applied for characterization of enantiomers. For this purpose, we use a special chiral sample. This is a dielectric disk with a chiral-structure metal coating. Fig. 21 (*a*) shows the sample structure. Fig 21 (*b*) represents the helicity-parameter distribution in a dielectric disk ($\varepsilon_r = 30$) when magnetic bias is up. This is the distribution with evident lack of azimuth symmetry. On a chiral-structure metal coating the helicity parameter is zero. When this sample is placed on a surface of a ferrite disk, one obtains evident distinction of the MDM spectra at different orientations of a bias magnetic field. This is illustrated in Fig. 22 for a transmission coefficient of a microstrip structure with a thin-film ferrite disk.

## IV. Conclusion

Small quasi-2D ferrite disks with the magnetic-dipolar-oscillation spectra are sources of peculiar microwave near fields. These fields, studied analytically in the preceding paper [5], are characterized by topologically distinctive power-flow vortices, non-zero helicity, and a torsion degree of freedom. The numerical and experimental studies shown in the present paper confirm main statements of the theory in Ref. [5].

Among a series of interesting properties of the microwave near fields, originated from quasi-2D ferrite disks with the magnetic-dipolar-oscillation spectra, there are interactions of such near fields with matter. Transformation of the MDM spectrum due to dielectric samples abutting to the surface of a ferrite disk was observed experimentally, for the first time, in Ref. [7]. In the present studies, we showed that the transformation of the MDM spectrum due to dielectric samples is strongly related to the helicity properties of the MDM near fields. We also showed that in virtue of the near-field helicity one can effectively observe at microwaves the enantiomeric properties of the samples. Use of subwavelength MDM fields with energy localization and symmetry breakings opens a perspective for unique microwave applications. Presently, the precise spectroscopic analysis of natural and artificial chiral structures is considered as one of the very important aspects in material characterization. Because of the helicity structure of the MDM near-fields, one can predict carrying out a precise spectroscopic analysis of natural and artificial chiral structures at microwaves.

**Figure captions**

Fig. 1. Frequency characteristics of a module of the reflection coefficient for a rectangular waveguide with an enclosed thin-film ferrite disk. The resonance modes are designated in succession by numbers n = 1, 2, 3… The coalescent resonances are denoted by single and double primes. An insert shows geometry of a structure.

Fig. 2. The helicity parameter for the 1$^{st}$ MDM (a ferrite disk is inside a rectangular waveguide).

Fig. 3. The helicity parameter for the 2$^{nd}$ (the resonance 2″) MDM (a ferrite disk is inside a rectangular waveguide).

Fig. 4. The helicity parameter for non-resonance frequencies (a ferrite disk is inside a rectangular waveguide).

Fig. 5. Magnetization distributions in a ferrite disk at the MDM resonances. (*a*) and (*b*) numerical results for the 1$^{st}$ and 2$^{nd}$ (the resonance 2″) MDMs, respectively; (*c*) and (*d*) analytical results for the 1$^{st}$ and 2$^{nd}$ MDMs, respectively, obtained based on the models in Refs. [8, 9]. The distributions are shown for a certain time phase.

Fig. 6. A sample of a ferrite disk with two loading dielectric cylinders placed inside a $TE_{10}$-mode rectangular waveguide.

Fig. 7. Frequency characteristics of a module of the reflection coefficient for the 1$^{st}$ MDM at different parameters of a symmetrical dielectric loading. Frequency $f_H = 8,456$ GHz is the Larmor frequency of an unloaded ferrite disk.

Fig. 8. Poynting vector distributions above a ferrite disk (on the plane parallel to the ferrite-disk plane and at distance 75 microns above a disk). The frequencies correspond to the resonance 1 of an unloaded (without dielectric cylinders) ferrite disk and the resonances 1″ of a ferrite disk with dielectric loadings.

Fig. 9. Numerically calculated helicity-parameter distributions for the 1$^{st}$ MDM at different dielectric constants of loading cylinders. The distributions are shown on the cross-section plane which passes through the diameter and the axis of the ferrite disk.

Fig. 10. The helicity parameter distributions for the 1$^{st}$ MDM at different dielectric constants of loading cylinders. The cross-section planes are parallel to the ferrite-disk plane and are at different distances from the ferrite surface: (a) 25 mkm, (b) 75 mkm, (c) 150 mkm.

Fig. 11. Space angle between vectors $\vec{E}$ and $\vec{\nabla} \times \vec{E}$ for the 1$^{st}$ MDM at different dielectric constants of loading cylinders. (*a*) $\varepsilon_r = 1$; (*b*) $\varepsilon_r = 30$; (*c*) $\varepsilon_r = 50$. Points *A* and *B* are singular points in dielectrics, where the helicity of the near fields changes its sign.

Fig. 12. The electric and magnetic fields (at a certain time phase) in a structure of a ferrite disk with loading dielectrics. (*a*) and (*b*) the electric and magnetic fields inside a ferrite disk for dielectric loadings of $\varepsilon_r = 1$ and $\varepsilon_r = 50$, respectively; (*c*) the electric and magnetic fields in a dielectric ($\varepsilon_r = 50$) on a plane 0.75 mm above a surface of a ferrite disk (the plane is below a singular point *A*); (*d*) the electric and magnetic fields in a dielectric ($\varepsilon_r = 50$) on a plane 1.7 mm above a surface of a ferrite disk (the plane is above a singular point *A*).

Fig. 13. A sample of a ferrite disk with one loading dielectric cylinder placed inside a $TE_{10}$-mode rectangular waveguide.

Fig. 14. Frequency characteristics of a module of the reflection coefficient for the 1$^{st}$ MDM at different parameters of a dielectric cylinder.

Fig. 15. Numerically calculated helicity parameter distributions for the 1$^{st}$ MDM at different dielectric constants of a loading cylinder. The distributions are shown on the cross-section plane which passes through the diameter and the axis of the ferrite disk.

Fig. 16. Frequency characteristic of a module of the transmission coefficient for a microstrip structure with a thin-film ferrite disk. An insert shows geometry of a structure.

Fig. 17. The helicity parameter for a microstrip structure with a ferrite disk. (a) For the 1$^{st}$ radial-mode resonance frequency; (b) at non-resonance frequencies.

Fig. 18. Transformation of the MDM spectrum due to a dielectric loading in a microstrip structure (numerical results).

Fig. 19. Transformation of the MDM spectrum due to a dielectric loading in a microstrip structure (experimental results).

Fig. 20. The helicity parameter for the 1$^{st}$ radial mode at opposite directions of a normal bias magnetic field. A ferrite disk is placed in a microstrip structure without a loading dielectric. (*a*) Magnetic bias is up; (*b*) magnetic bias is down.

Fig. 21. A chiral sample. (*a*) Sample structure; (*b*) the helicity parameter distribution in a dielectric disk when magnetic bias is up.

Fig. 22. Measuring of chirality with use of opposite directions of a DC magnetic field (numerical results). An insert shows the position of a chiral sample in a microstrip structure.

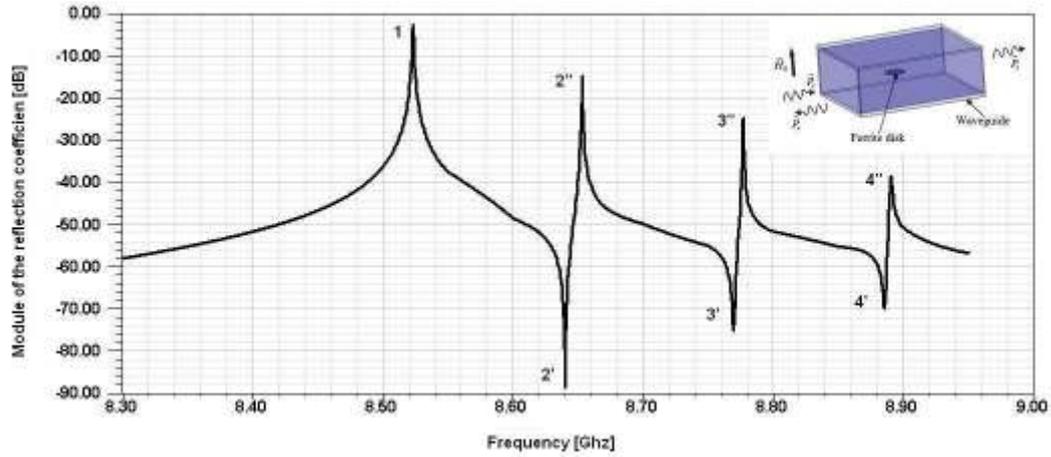

Fig. 1. Frequency characteristics of a module of the reflection coefficient for a rectangular waveguide with an enclosed thin-film ferrite disk. The resonance modes are designated in succession by numbers n = 1, 2, 3… The coalescent resonances are denoted by single and double primes. An insert shows geometry of a structure.

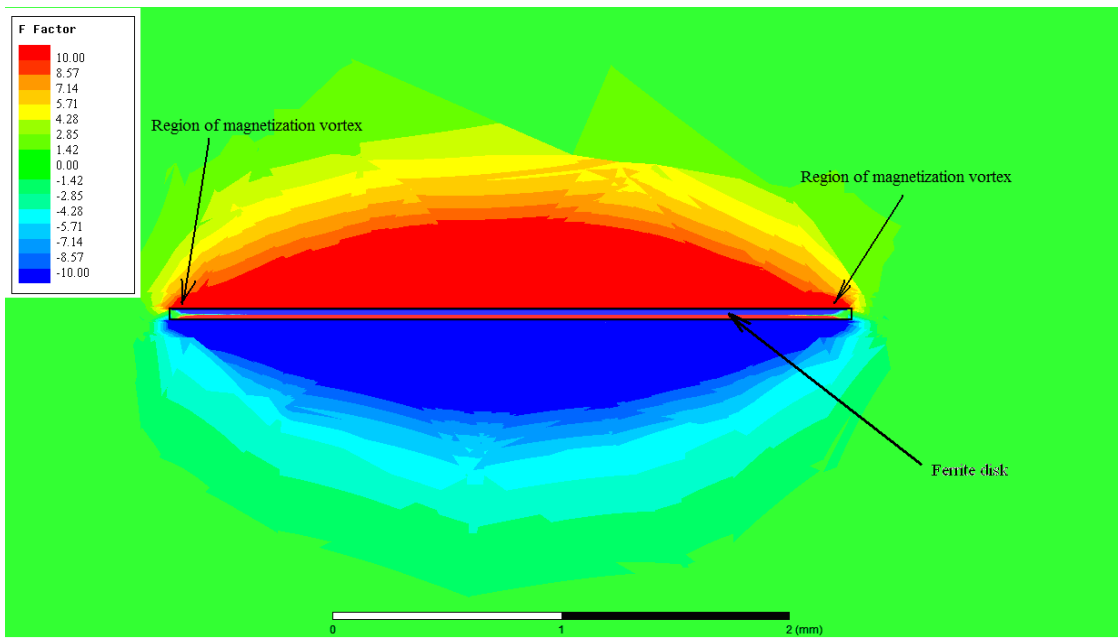

Fig. 2. The helicity parameter for the 1$^{st}$ MDM (a ferrite disk is inside a rectangular waveguide).

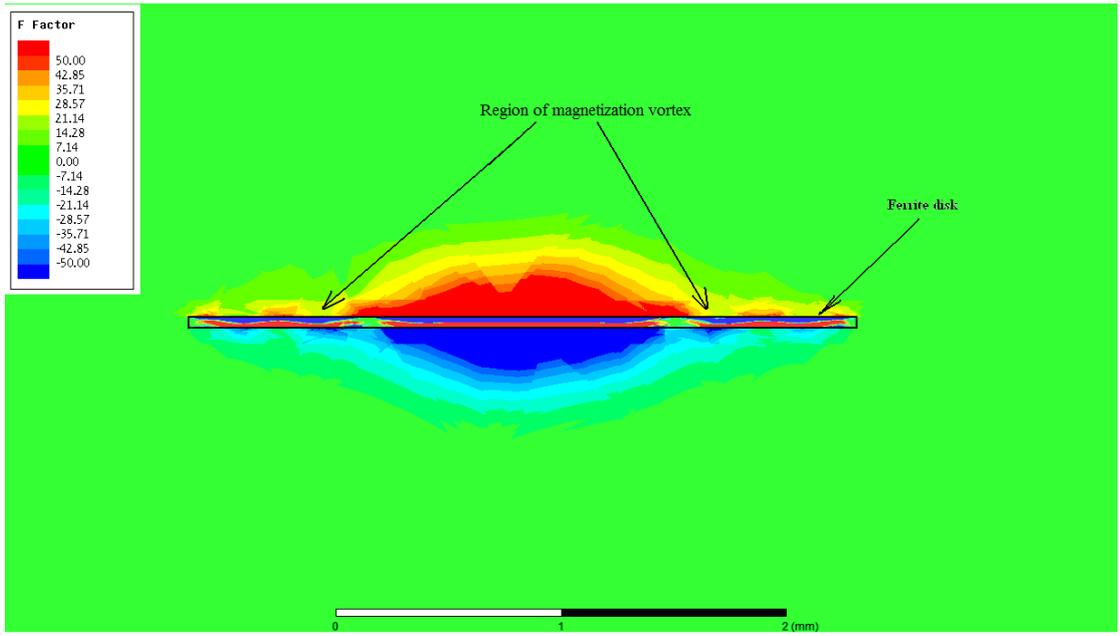

Fig. 3. The helicity parameter for the 2$^{nd}$ (the resonance 2″) MDM (a ferrite disk is inside a rectangular waveguide).

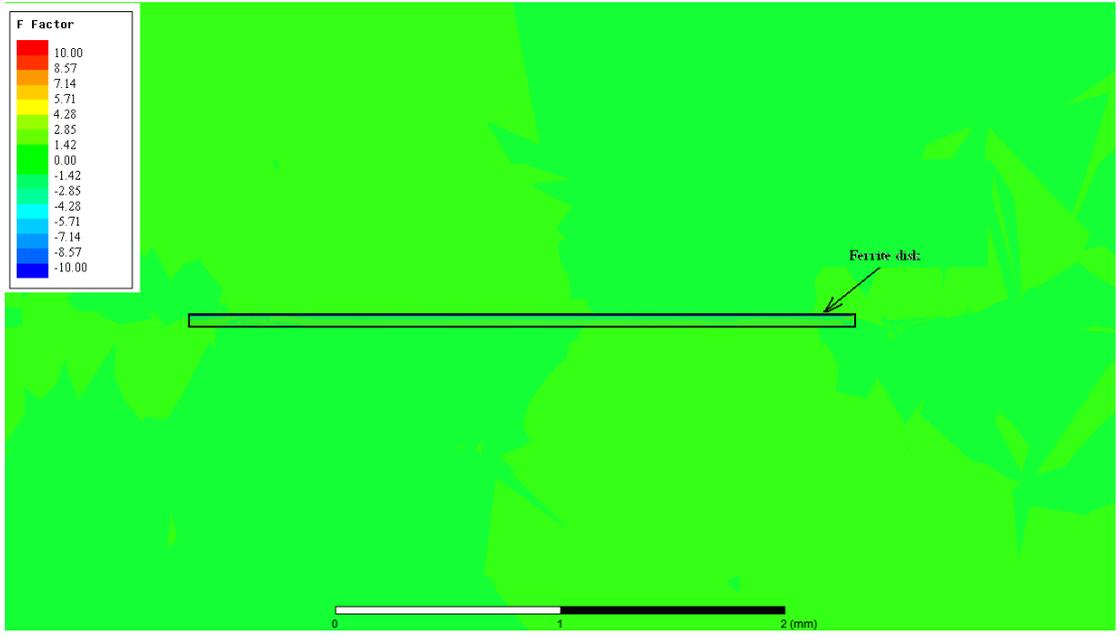

Fig. 4. The helicity parameter for non-resonance frequencies (a ferrite disk is inside a rectangular waveguide).

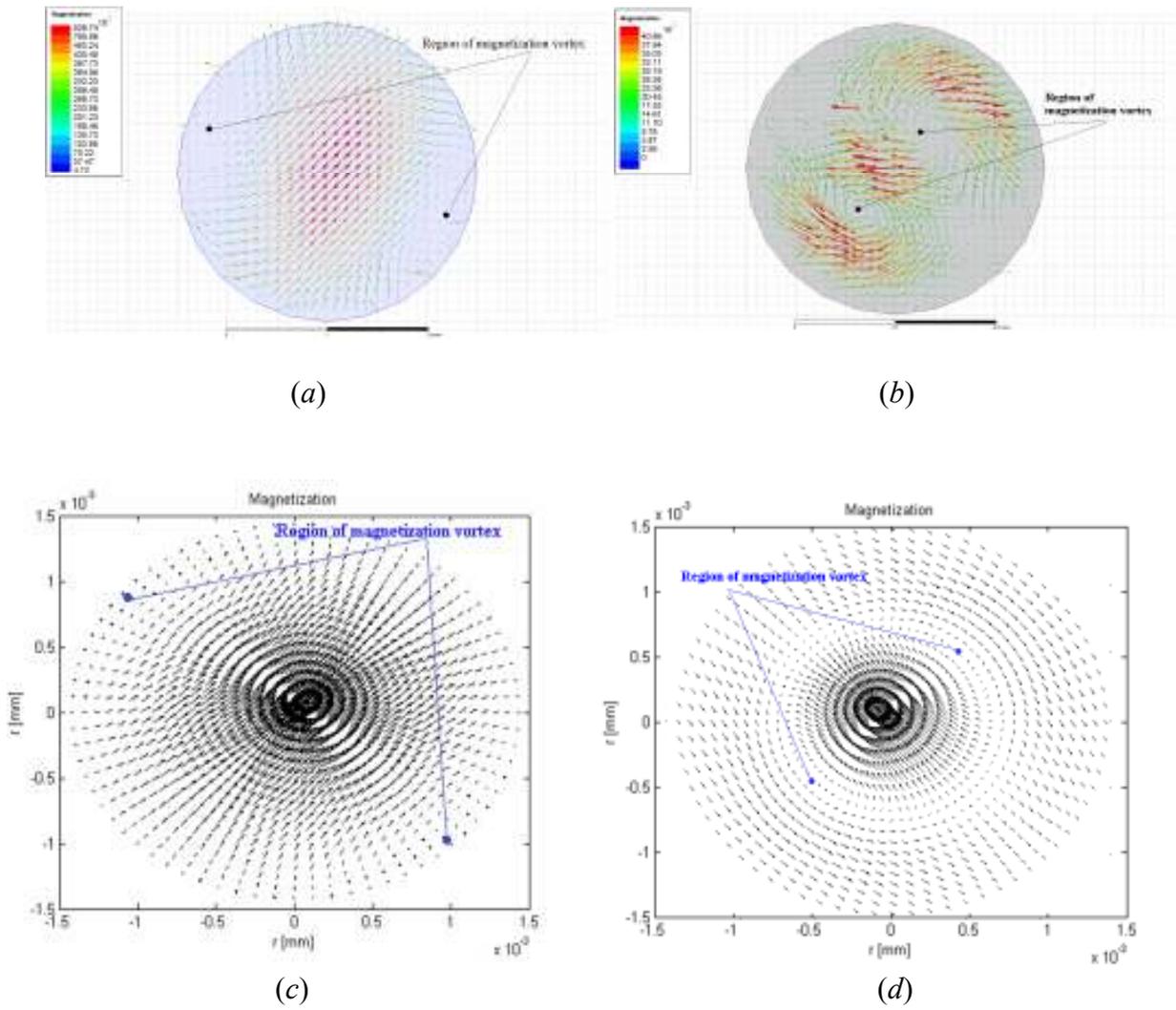

Fig. 5. Magnetization distributions in a ferrite disk at the MDM resonances. (*a*) and (*b*) numerical results for the 1$^{st}$ and 2$^{nd}$ (the resonance 2") MDMs, respectively; (*c*) and (*d*) analytical results for the 1$^{st}$ and 2$^{nd}$ MDMs, respectively, obtained based on the models in Refs. [8, 9]. The distributions are shown for a certain time phase.

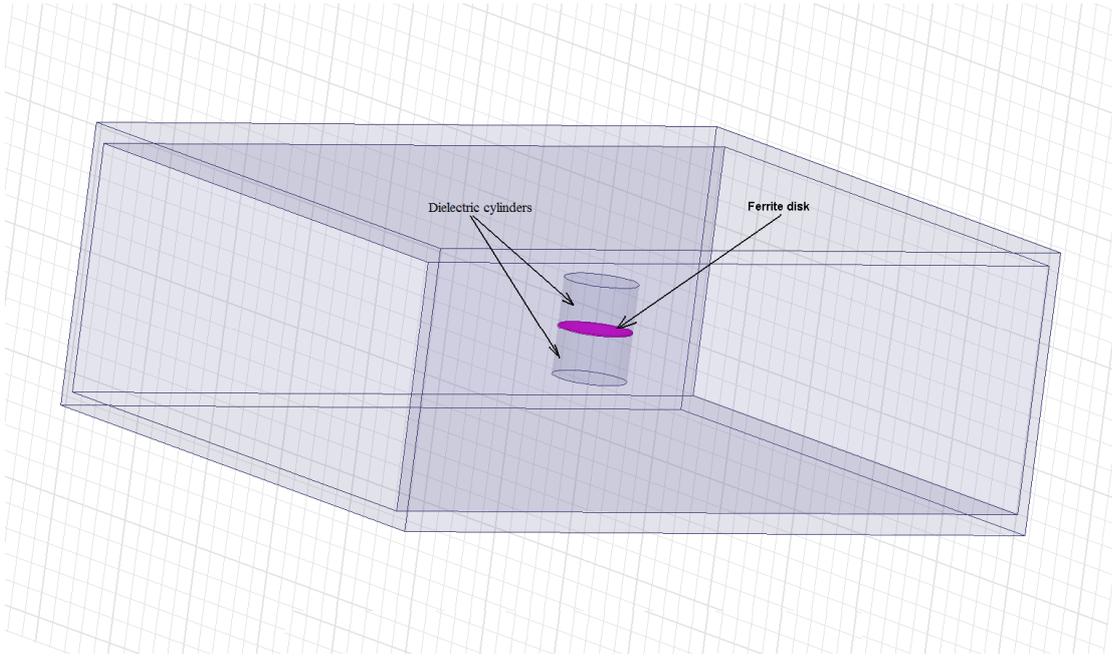

Fig. 6. A sample of a ferrite disk with two loading dielectric cylinders placed inside a $TE_{10}$-mode rectangular waveguide.

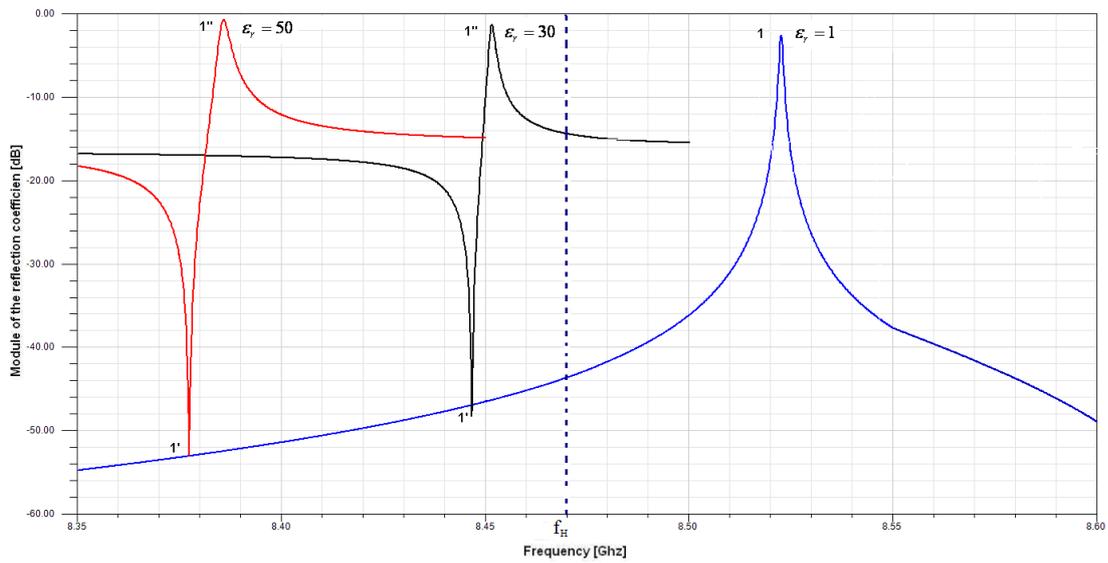

Fig. 7. Frequency characteristics of a module of the reflection coefficient for the 1st MDM at different parameters of a symmetrical dielectric loading. Frequency $f_H = 8,456$ GHz is the Larmor frequency of an unloaded ferrite disk.

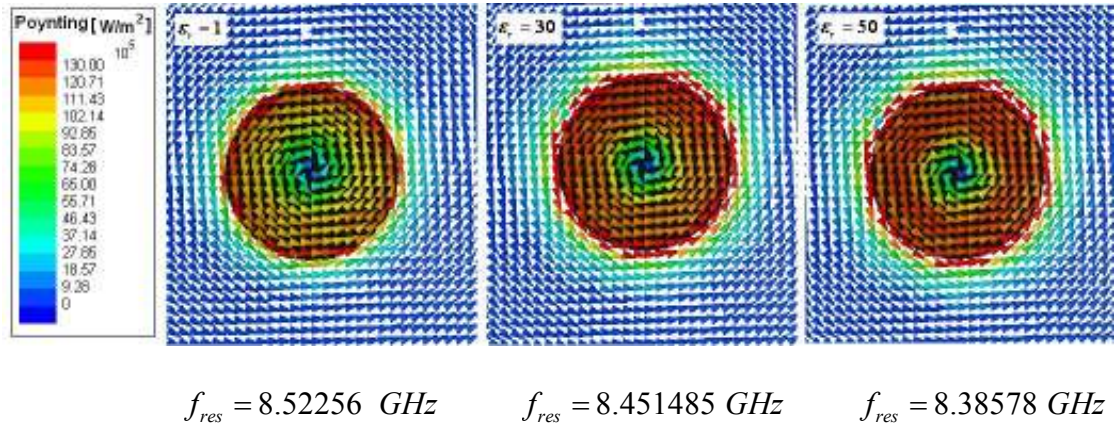

$f_{res} = 8.52256\ GHz$        $f_{res} = 8.451485\ GHz$        $f_{res} = 8.38578\ GHz$

Fig. 8. Poynting vector distributions above a ferrite disk (on the plane parallel to the ferrite-disk plane and at distance 75 microns above a disk). The frequencies correspond to the resonance 1 of an unloaded (without dielectric cylinders) ferrite disk and the resonances 1″ of a ferrite disk with dielectric loadings.

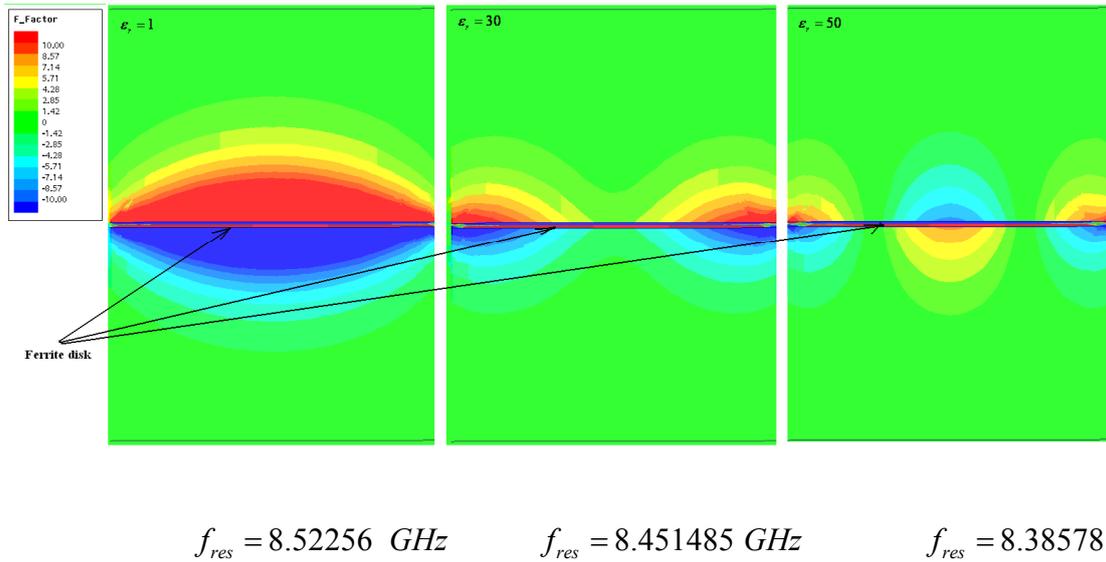

$f_{res} = 8.52256\ GHz$        $f_{res} = 8.451485\ GHz$        $f_{res} = 8.38578\ GHz$

Fig. 9. Numerically calculated helicity-parameter distributions for the 1$^{st}$ MDM at different dielectric constants of loading cylinders. The distributions are shown on the cross-section plane which passes through the diameter and the axis of the ferrite disk.

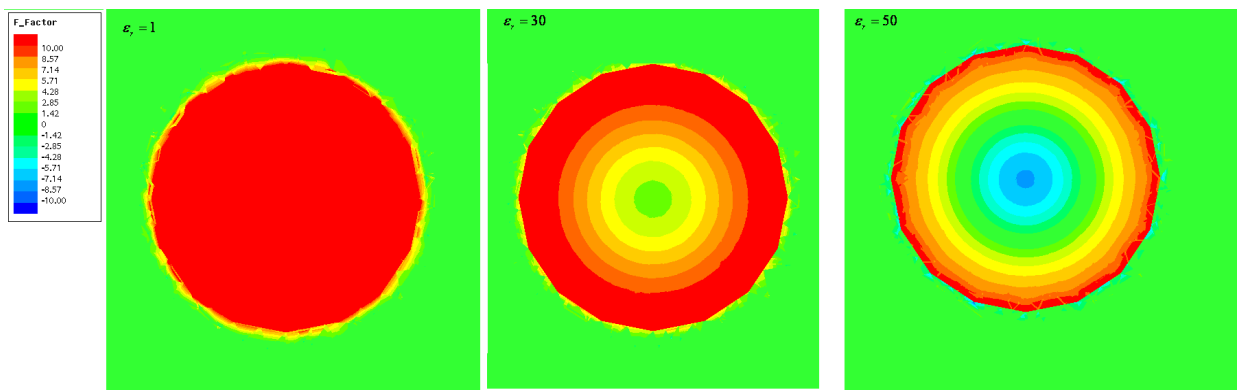

$f_{res} = 8.52256\ GHz$     $f_{res} = 8.451485\ GHz$     $f_{res} = 8.38578\ GHz$

(a)

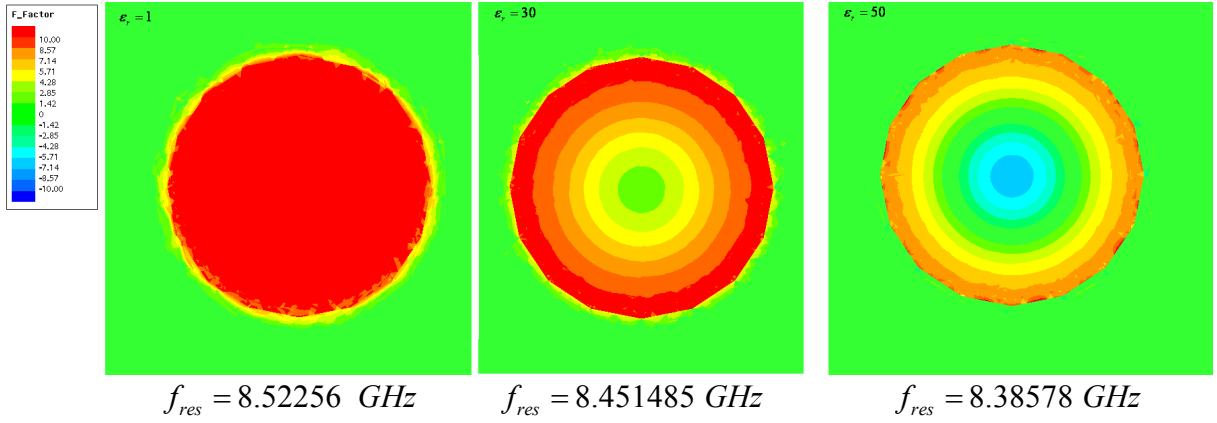

$f_{res} = 8.52256\ GHz$     $f_{res} = 8.451485\ GHz$     $f_{res} = 8.38578\ GHz$

(b)

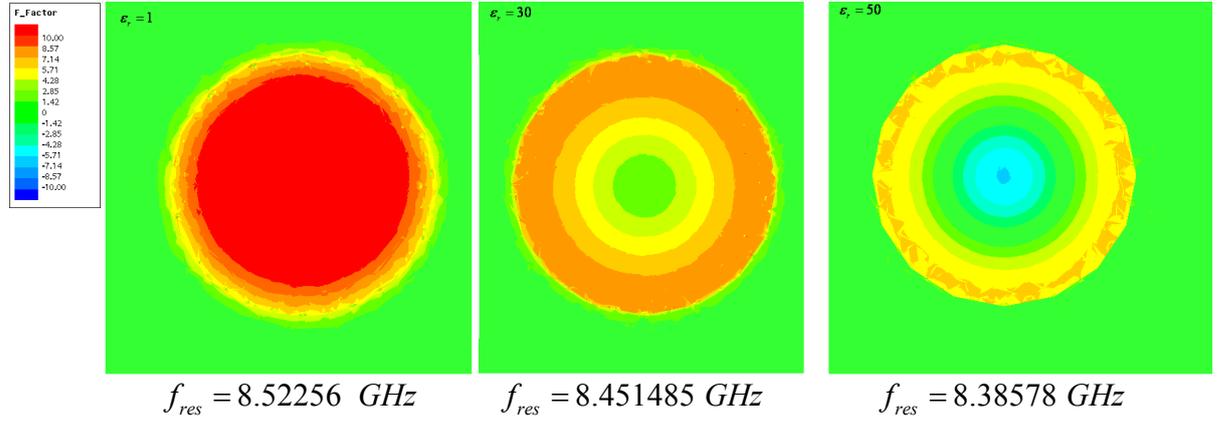

$f_{res} = 8.52256\ GHz$     $f_{res} = 8.451485\ GHz$     $f_{res} = 8.38578\ GHz$

(c)

Fig. 10. The helicity parameter distributions for the 1$^{st}$ MDM at different dielectric constants of loading cylinders. The cross-section planes are parallel to the ferrite-disk plane and are at different distances from the ferrite surface: (a) 25 mkm, (b) 75 mkm, (c) 150 mkm.

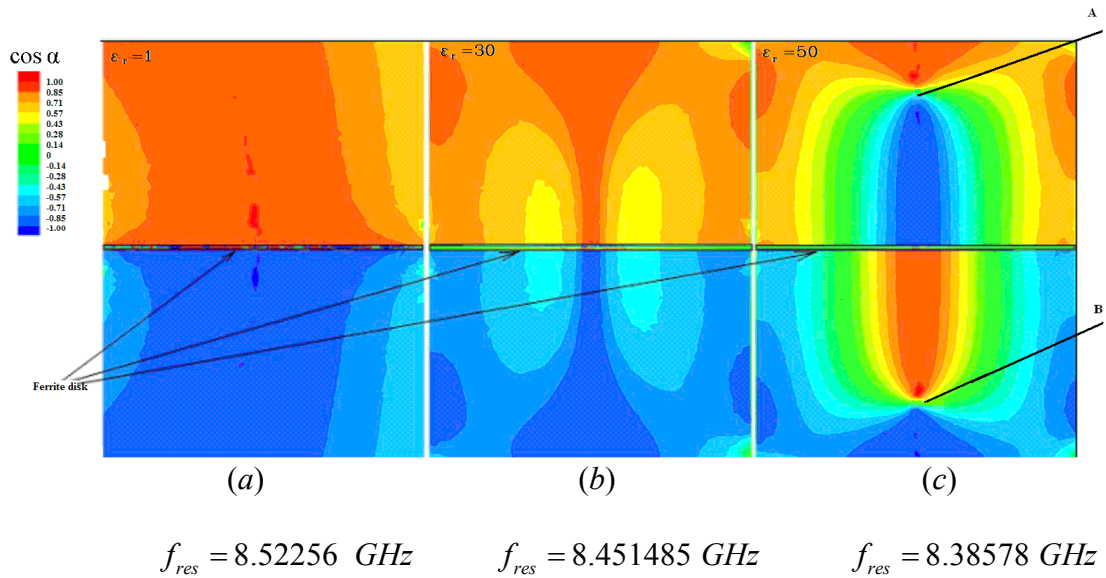

$f_{res} = 8.52256\ GHz$       $f_{res} = 8.451485\ GHz$       $f_{res} = 8.38578\ GHz$

Fig. 11. Space angle between vectors $\vec{E}$ and $\vec{\nabla} \times \vec{E}$ for the 1$^{st}$ MDM at different dielectric constants of loading cylinders. (*a*) $\varepsilon_r = 1$; (*b*) $\varepsilon_r = 30$; (*c*) $\varepsilon_r = 50$. Points *A* and *B* are singular points in dielectrics, where the helicity of the near fields changes its sign.

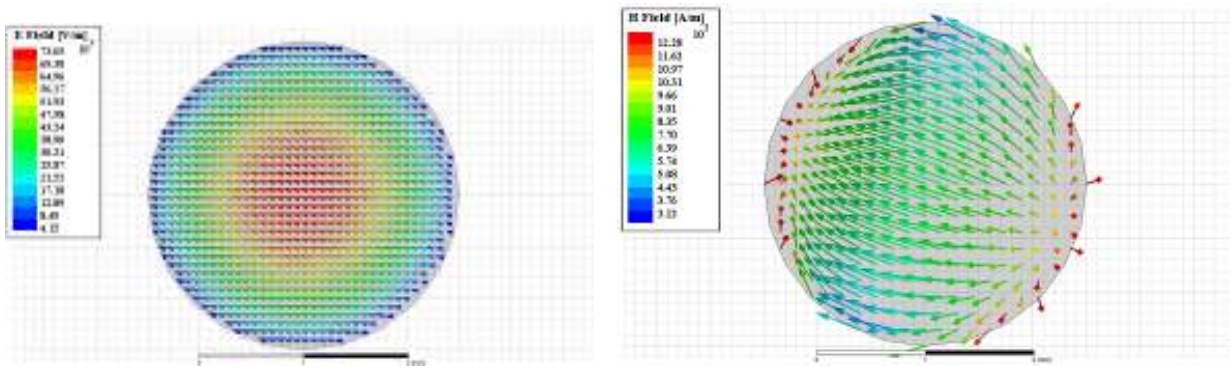

(*a*)

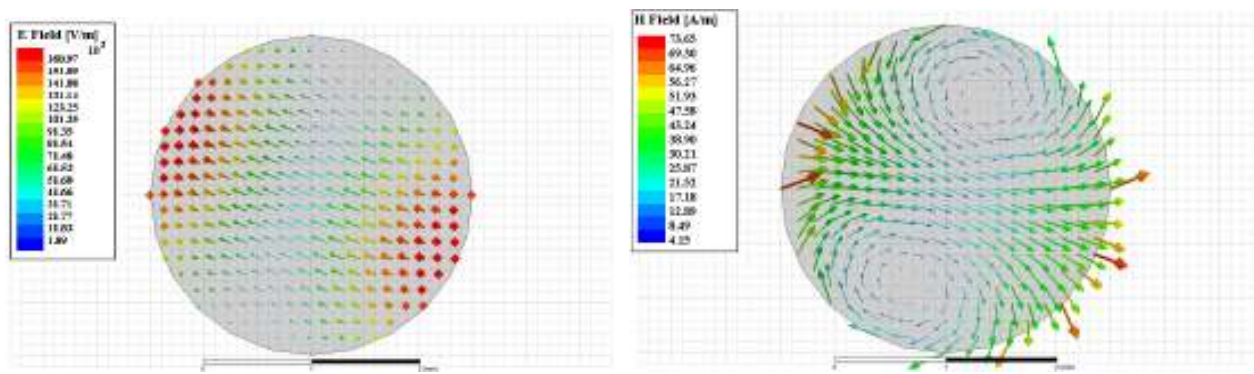

(*b*)

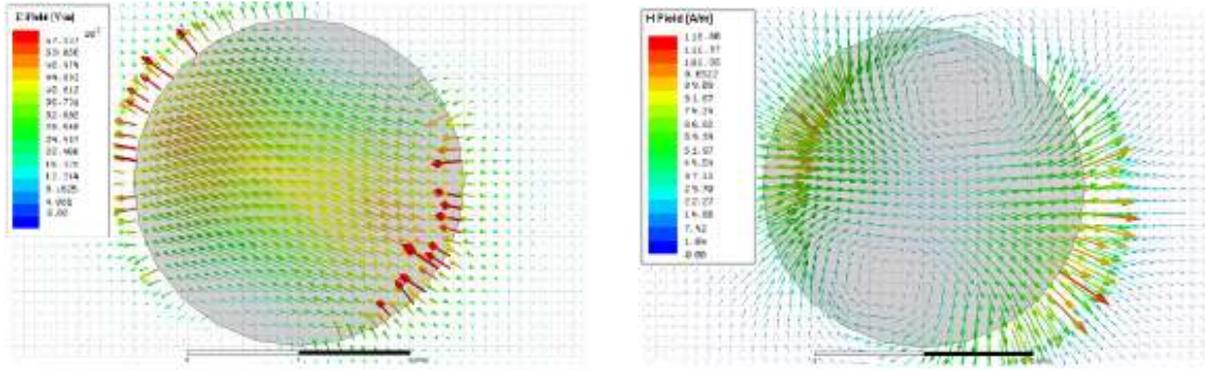

(*c*)

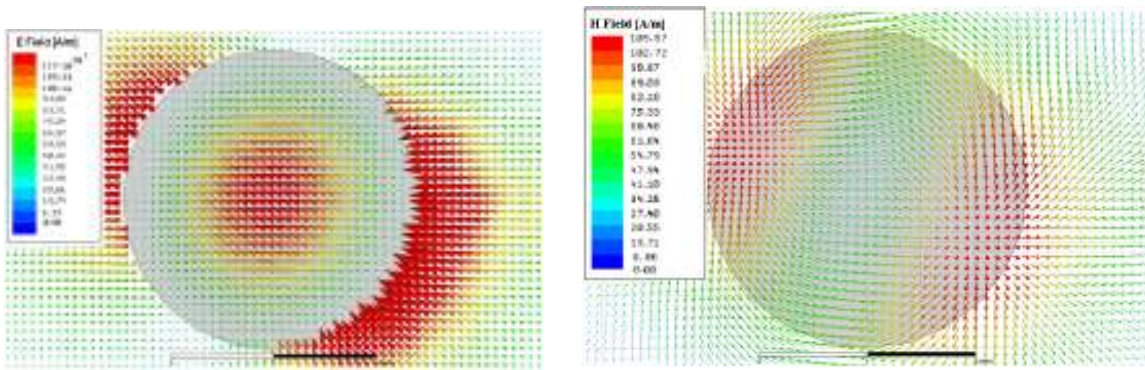

(*d*)

Fig. 12. The electric and magnetic fields (at a certain time phase) in a structure of a ferrite disk with loading dielectrics. (*a*) and (*b*) the electric and magnetic fields inside a ferrite disk for dielectric loadings of $\varepsilon_r = 1$ and $\varepsilon_r = 50$, respectively; (*c*) the electric and magnetic fields in a dielectric ($\varepsilon_r = 50$) on a plane 0.75 mm above a surface of a ferrite disk (the plane is below a singular point *A*); (*d*) the electric and magnetic fields in a dielectric ($\varepsilon_r = 50$) on a plane 1.7 mm above a surface of a ferrite disk (the plane is above a singular point *A*).

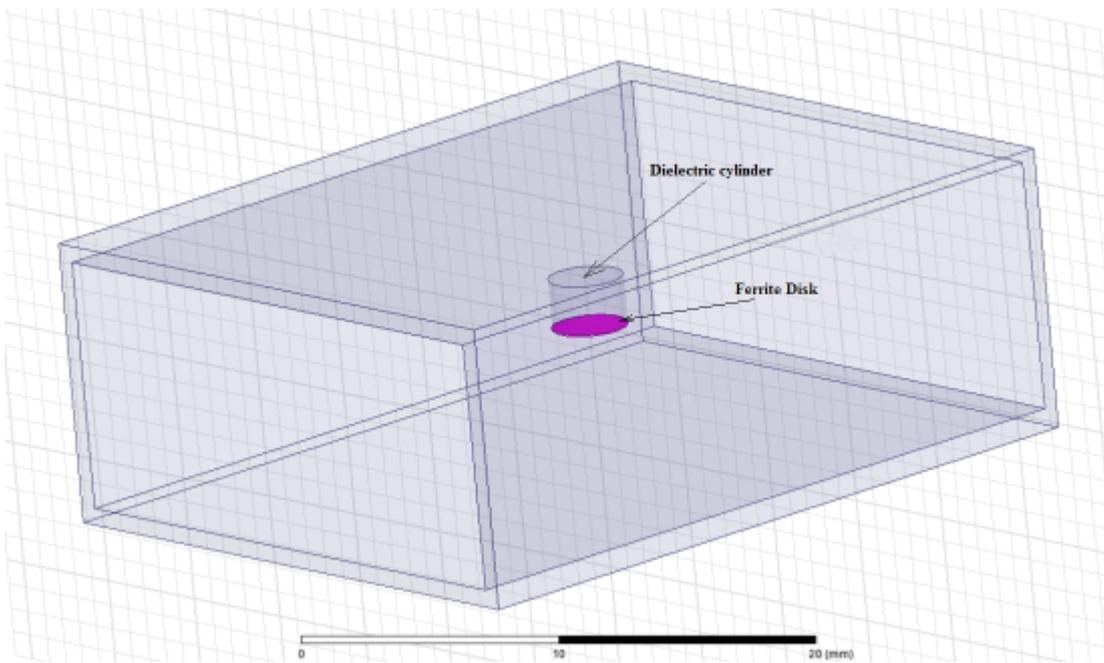

Fig. 13. A sample of a ferrite disk with one loading dielectric cylinder placed inside a $TE_{10}$-mode rectangular waveguide.

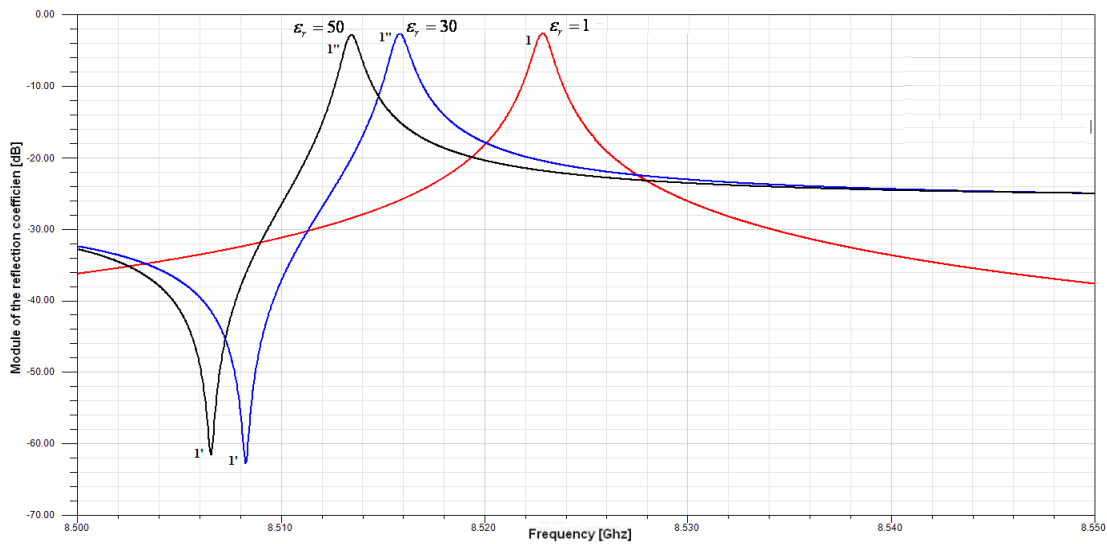

Fig. 14. Frequency characteristics of a module of the reflection coefficient for the 1st MDM at different parameters of a dielectric cylinder.

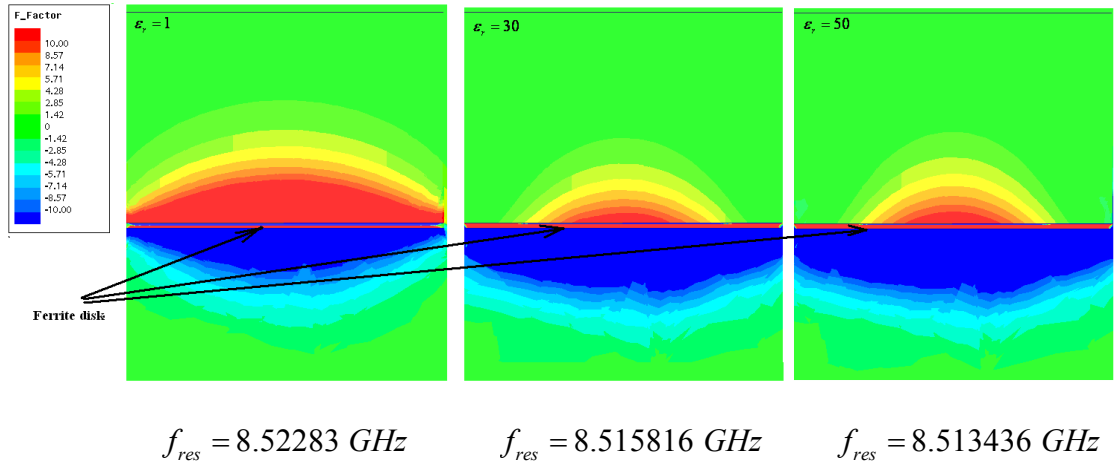

$f_{res} = 8.52283\ GHz$　　　$f_{res} = 8.515816\ GHz$　　　$f_{res} = 8.513436\ GHz$

Fig. 15. Numerically calculated helicity parameter distributions for the 1$^{st}$ MDM at different dielectric constants of a loading cylinder. The distributions are shown on the cross-section plane which passes through the diameter and the axis of the ferrite disk.

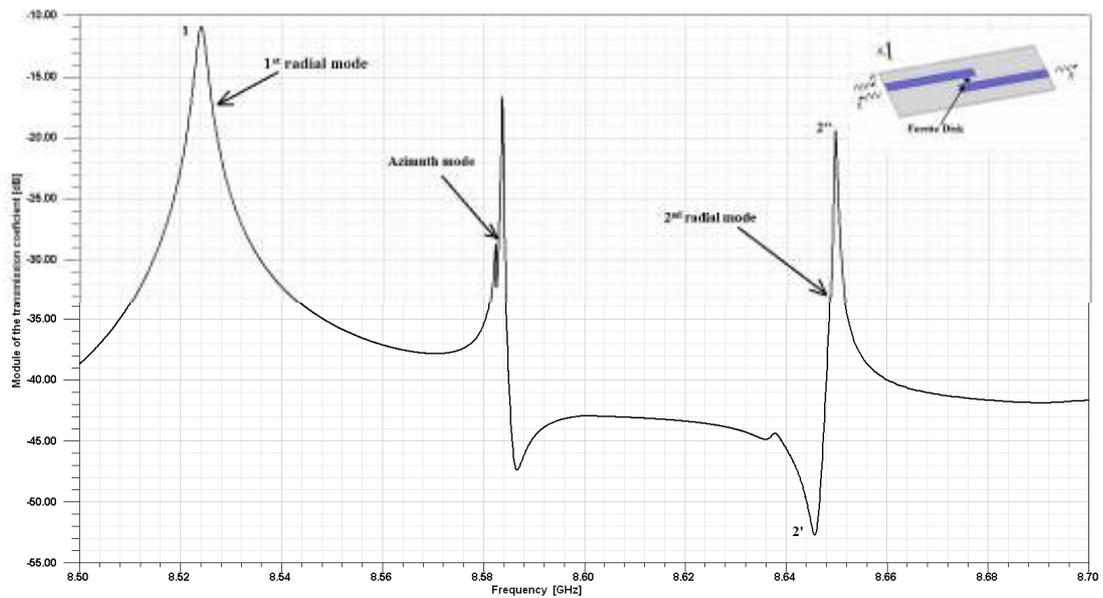

Fig. 16. Frequency characteristic of a module of the transmission coefficient for a microstrip structure with a thin-film ferrite disk. An insert shows geometry of a structure.

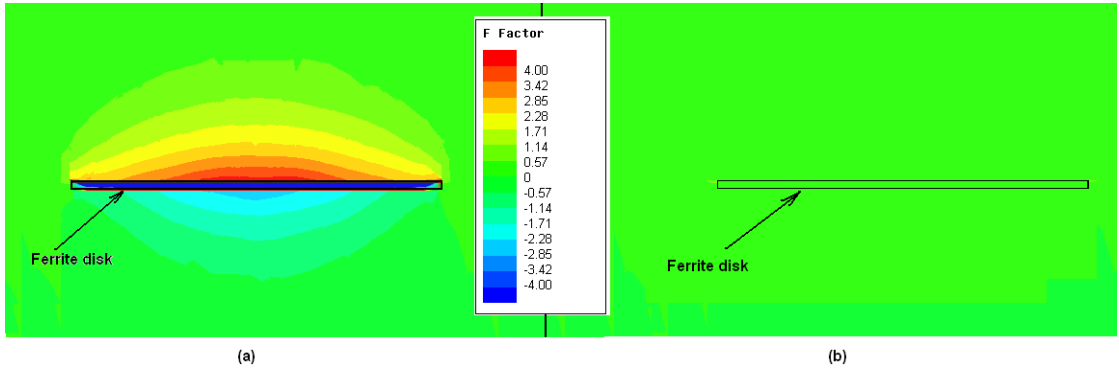

Fig. 17. The helicity parameter for a microstrip structure with a ferrite disk. (a) For the 1st radial-mode resonance frequency; (b) at non-resonance frequencies.

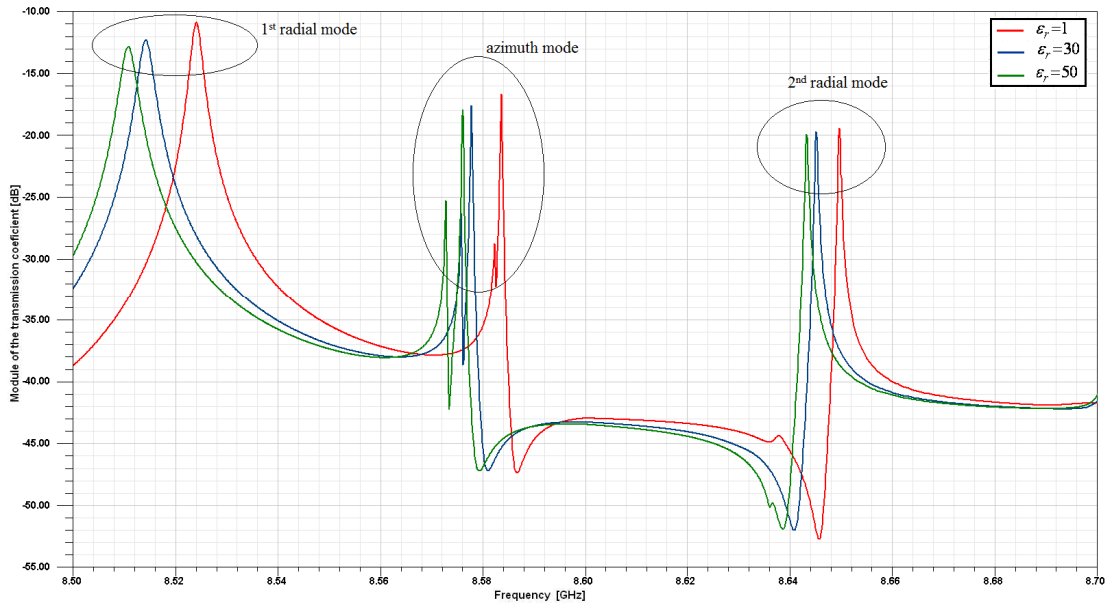

Fig. 18. Transformation of the MDM spectrum due to a dielectric loading in a microstrip structure (numerical results).

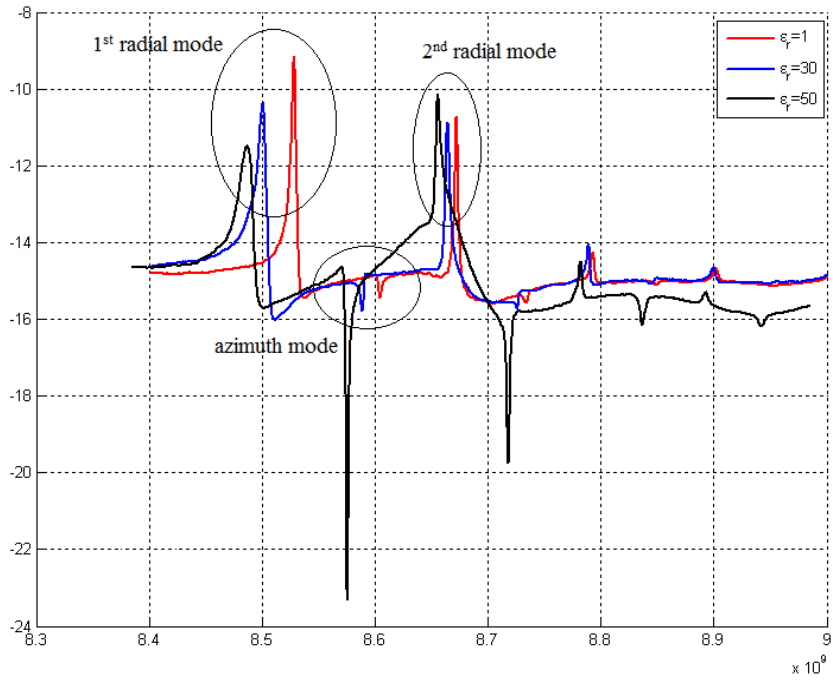

Fig. 19. Transformation of the MDM spectrum due to a dielectric loading in a microstrip structure (experimental results).

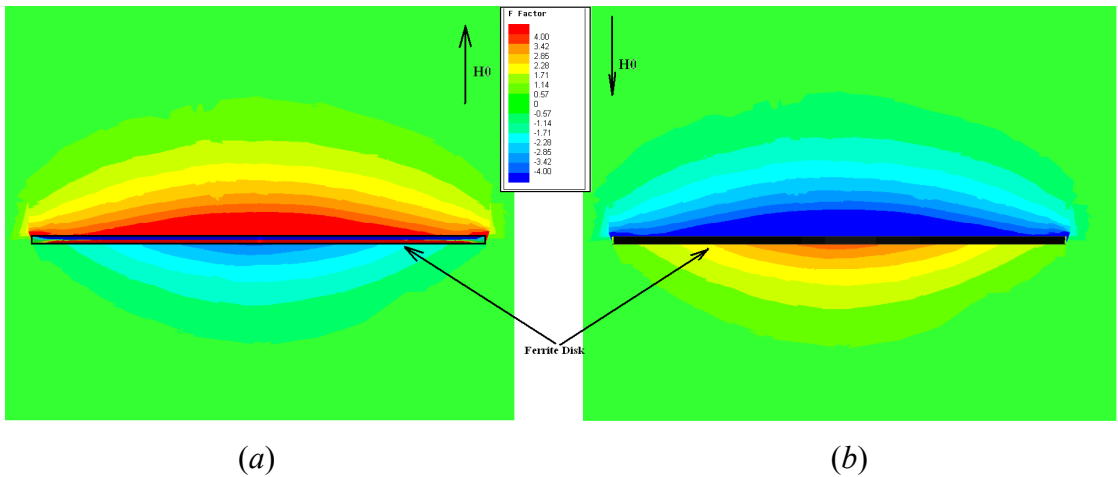

(*a*)            (*b*)

Fig. 20. The helicity parameter for the 1$^{st}$ radial mode at opposite directions of a normal bias magnetic field. A ferrite disk is placed in a microstrip structure without a loading dielectric. (*a*) Magnetic bias is up; (*b*) magnetic bias is down.

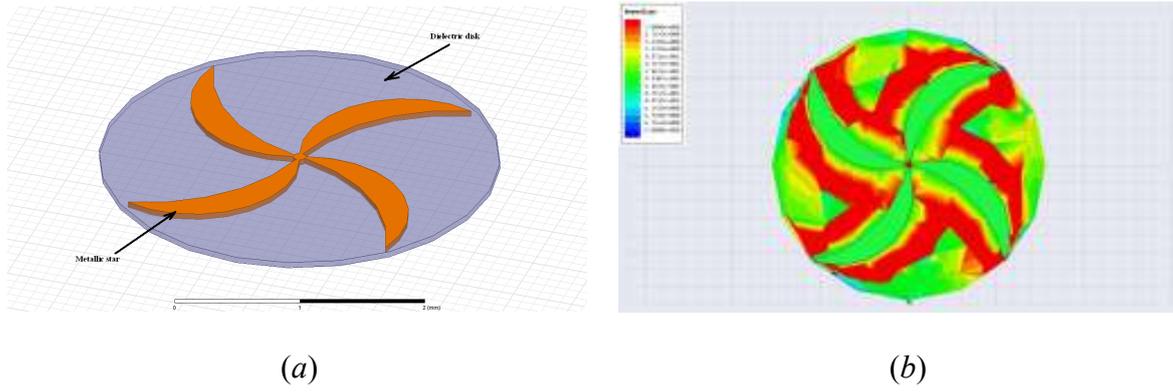

Fig. 21. A chiral sample. (*a*) Sample structure; (*b*) the helicity parameter distribution in a dielectric disk when magnetic bias is up.

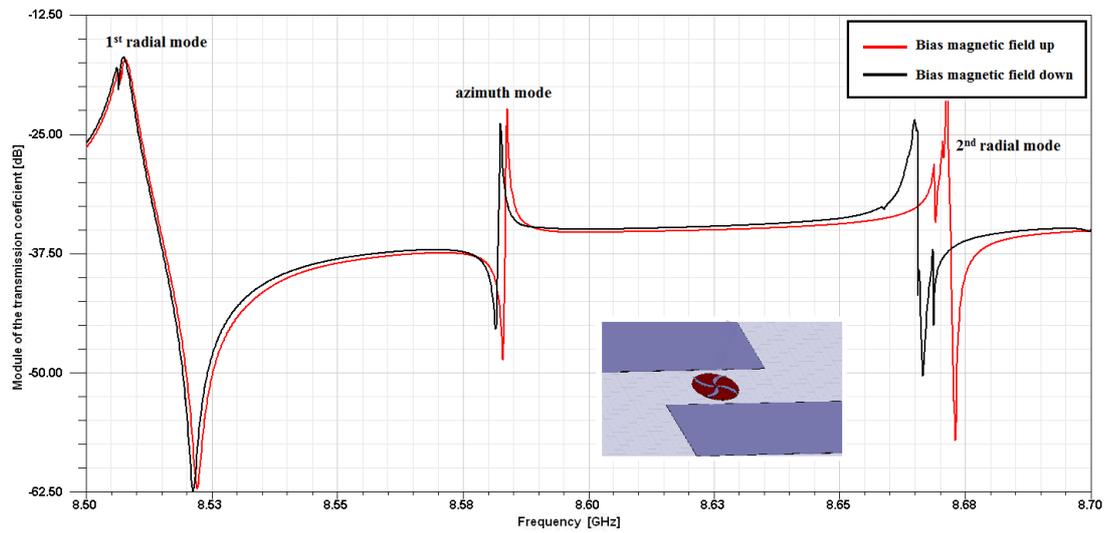

Fig. 22. Measuring of chirality with use of opposite directions of a DC magnetic field (numerical results). An insert shows the position of a chiral sample in a microstrip structure.

23